\documentclass[a4paper,11pt]{article}
\pdfoutput=1 
\usepackage{jinstpub} 
\usepackage{subfigure}
\usepackage{amssymb}
\usepackage{amsthm}
\usepackage{amsmath}
\usepackage[figuresright]{rotating}
\usepackage{graphicx}
\usepackage{tikz}			
\usetikzlibrary{decorations.pathmorphing}       
\usetikzlibrary{decorations.pathreplacing}      
\usetikzlibrary{positioning}                    
\usetikzlibrary{arrows}                         
\usetikzlibrary{patterns}                       
\usetikzlibrary{shapes}                         
\usetikzlibrary{scopes}
\usepackage{textcomp}
\usepackage[T1]{fontenc}
\usepackage[tight]{units}
\usepackage{hyperref}
\usepackage{lineno}

\begin{document}
\title{Results from the first cryogenic NaI detector for the COSINUS project}

\author[a]{G.~Angloher,}

\author[b,c]{P.~Carniti,} 

 \author[b,c]{L.~Cassina,} 

 \author[b,c]{L.~Gironi,}

\author[b,c]{C.~Gotti,}

\author[d]{A.~G\"utlein,}

\author[b]{M.~Maino,}

\author[a]{M.~Mancuso,}

\author[e]{L.~Pagnanini,}

\author[b]{G.~Pessina,}

\author[a]{F.~Petricca,}

 \author[f]{S.~Pirro,}
 
 \author[a]{F.~Pr\"obst,}
 
 \author[d]{R.~Puig,} 
 
 \author[a,1]{F.~Reindl \note{corresponding author, present address: INFN, Sezione di Roma 1, I-00185 Roma - Italy},}
\emailAdd{florian.reindl@mpp.mpg.de}

\author[e,2]{K.~Sch\"affner \note{corresponding author},}
\emailAdd{karoline.schaeffner@lngs.infn.it}

\author[d]{J.~Schieck,}
 
\author[a,3]{W.~Seidel \note{deceased 19/02/2017}}

\affiliation[a]{Max-Planck-Institut f\"ur Physik, D-80805 M\"unchen - Germany}
\affiliation[b]{INFN - Sezione di Milano Bicocca, I-20126  Milano - Italy}
\affiliation[c]{Dipartimento di Fisica, Universit\`{a} di Milano-Bicocca, I-20126 Milano - Italy}
\affiliation[d]{Institut f\"ur Hochenergiephysik der \"Osterreichischen Akademie der Wissenschaften, A-1050 Wien - Austria and Atominstitut, Vienna University of Technology, A-1020 Wien - Austria}
\affiliation[e]{Gran Sasso Science Institute, I-67100 L'Aquila - Italy}
\affiliation[f]{INFN - Laboratori Nazionali del Gran Sasso, I-67010 Assergi - Italy}

\arxivnumber{1705.11028} 
\collaboration{COSINUS collaboration}

\abstract{%
Recently there is a flourishing and notable interest in the crystalline scintillator material sodium iodide (NaI) as target for direct dark matter searches. This is mainly driven by the long-reigning contradicting situation in the dark matter sector: the positive evidence for the detection of a dark matter modulation signal claimed by the DAMA/LIBRA collaboration is (under so-called standard assumptions) inconsistent with the null-results reported by most of the other direct dark matter experiments. We present the results of a first prototype detector using a new experimental approach in comparison to \textit{conventional} single-channel NaI scintillation light detectors: a NaI crystal operated as a scintillating calorimeter at milli-Kelvin temperatures simultaneously providing a phonon (heat) plus scintillation light signal and particle discrimination on an event-by-event basis. We evaluate energy resolution, energy threshold and further performance parameters of this prototype detector developed within the COSINUS R\&D project.%
}
\maketitle
\section{Introduction}
\label{intro}
Dark matter (DM) is a main ingredient of the cosmos, contributing five times more to its energy density than the visible Universe. However, its nature is one of the most important, yet unsolved questions in today's physics. Back in 1998 the DAMA/NaI collaboration (now DAMA/LIBRA) published a status report on their first experimental data: therein they find a hint for a modulation signal which in period and phase matches the expectation for dark matter \cite{Bernabei1998}. Today, the positive evidence for a modulation signal is supported at $> 9 \sigma$ C.L.~in more than 13 annual cycles \cite{DAMA2013}.  In the standard scenario, however, the DAMA/LIBRA claim is inconsistent with the null-results of most of the other direct detection experiments \cite{cdex_collaboration_limits_2014,supercdms_collaboration_wimp-search_2015,SuperCDMS,damic_collaboration_search_2016,Hehn2016,CRESST_LISE,CRESST2,brown_extending_2012,DarkSide50,lux_collaboration_results_2017,pandax-ii_collaboration_dark_2016,
xenon_collaboration_xenon100_2016,pico_collaboration_dark_2016,pico_collaboration_improved_2016,lee_search_2014}.
Since a model-independent and unambiguous cross-check of the observed signal by DAMA/LIBRA is only possible by using the identical target material, a series of R\&D projects and experiments using sodium iodide (NaI) as target are presently in construction phase or being upgraded \cite{adhikari_understanding_2017, NaI_KIMS_200kg, SABRE2015, Amare2016, DMIce2017}.

Employing a dual channel readout of a NaI detector is a new attempt to the field: conventional NaI-based experiments in direct dark matter detection are operated as scintillation detectors, collecting \textit{only} the scintillation light emitted from a particle interaction by using photomultiplier tubes (PMTs).

COSINUS (Cryogenic Observatory for SIgnatures seen in Next-generation Underground Search\-es) aims to make a more profound investigation of the DAMA modulation signal using a novel experimental approach combining the DAMA/LIBRA detector material NaI (neglecting the 0.1\% of Tl-dopant present in DAMA/LIBRA crystals) with a cryogenic detection technique \cite{angloher_cosinus_2016}. Operating NaI crystals as low-temperature scintillating calorimeters at about \unit[10]{mK} has two distinct advantages: a possibly lower energy threshold than DAMA/LIBRA, in particular for nuclear recoil events and the possibility for active background rejection by particle identification on an event-by-event basis. Such detectors have the potential to clarify the underlying nature of the DAMA/LIBRA modulation signal, in particular if it is a nuclear recoil signal or not.

In this paper we present the results regarding energy resolution and energy threshold from our first NaI scintillating calorimeter measurement carried out at the Laboratori Nazionali del Gran Sasso (LNGS) in the frame of the COSINUS project.

\section{Detector Design}
\label{sec:2}
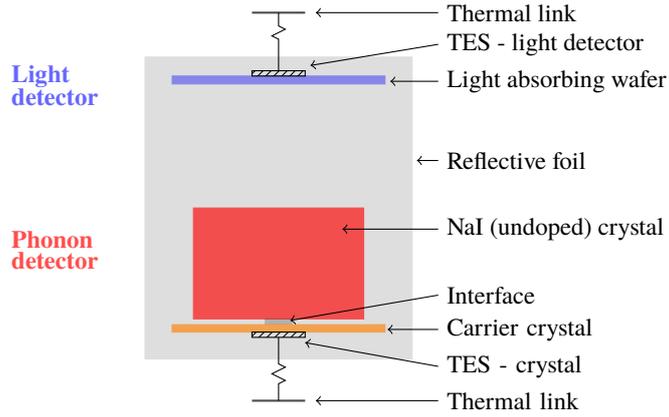
\begin{figure}
\centering
  \begin{tikzpicture}[scale=0.7, every node/.style={scale=0.85}]
   
    \draw[color=black, thick, fill=black, opacity=0.13]
      (-1.5,-2.85) rectangle (3.5,2.85);
    \draw [color=black, thick, opacity=0.7]
      (0.5,3.7) -- (1.5,3.7);
    \draw[decorate, decoration={zigzag, pre length=5, post length=10, segment length=5}]
      (1,3.7) -- (1,2.6);
    \draw[color=blue, fill=blue, opacity=0.4]
      (-1.0,2.5) rectangle (3.0,2.35);
    \draw [color=black, pattern=north east lines]
    (0.5,2.6) rectangle (1.5,2.5);
    \draw[color=red, fill=red, opacity=0.65]
      (-.60,0.0) rectangle (2.6,-2.1);
    \draw[color=black, fill=black, opacity=0.2]
    (0.75,-2.1) rectangle (1.25,-2.2);
    \draw[color=orange, fill=orange, opacity=0.65]
      (-1,-2.35) rectangle (3,-2.21);
    \draw[decorate, decoration={zigzag, pre length=10, post length=5, segment length=5}]
      (1,-2.45) -- (1,-3.65);
    \draw[color=black, thick, opacity=0.7]
      (0.5,-3.65) -- (1.5,-3.65);
    \draw[color=black, pattern=north east lines]
    (0.5,-2.35) rectangle (1.5,-2.45);
    \draw[color=black,->] (4.0,3.7) node[color=black, right]{Thermal link} -- (1.7, 3.7);
    \draw[color=black,->] (4.0,3.1) node[color=black, right]{TES - light detector} -- (1.6, 2.7);
    \draw[color=black,->] (4.0,2.4) node[color=black, right]{Light absorbing wafer} -- (3.1, 2.4);
    \draw[color=black,->] (4.0,0.9) node[color=black, right]{Reflective foil} -- (3.6, 0.9);
    \draw[color=black,->] (4.0,-0.40) node[color=black, right]{NaI (undoped) crystal} -- (2.2, -0.4);
    \draw[color=black,->](4.0,-3.0) node[color=black, text width=2.3cm, right]{TES - crystal} --(1.5, -2.55);    
    \draw[color=black,->](4.0,-1.65) node[color=black, text width=2.3cm, right]{Interface} --(1.22, -2.13);
    \draw[color=black,->](4.0,-2.3) node[color=black, text width=2.5cm, right]{Carrier crystal} --(3.05, -2.3);
    \draw[color=black,->](4.0,-3.65) node[color=black, text width=2.3cm, right]{Thermal link} -- (1.7, -3.65);
    \draw(-2.5,-0.6) node[color=red,thick, opacity=0.7, text width=2.5cm]{\textbf{Phonon}};
    \draw(-2.5,-1.0) node[color=red, thick, opacity=0.7, text width=2.5cm]{\textbf{detector}};
    \draw(-2.5,2.5) node[color=blue, thick, opacity=0.6, text width=2.5cm]{\textbf{Light}};
    \draw(-2.5,2.1) node[color=blue, thick, opacity=0.6, text width=2.5cm]{\textbf{detector}};
  \end{tikzpicture}
  \caption{Schema of the first COSINUS detector module consisting of an undoped NaI target crystal and a cryogenic light detector. Both detectors, operated at milli-Kelvin temperatures, are read out by transition edge sensors (TESs) and are surrounded by a reflective foil (LuMirror\texttrademark~lateral side, VM2002\texttrademark~Radiant Mirror Film above light detector).}
  \label{pic:module_schema}
\end{figure}

A cryogenic scintillating calorimeter detects two coincident signals that are both caused by a particle interaction: the phonon or heat signal indicates the energy deposited in the crystal by the interacting particle, whereas the scintillation light signal allows for particle identification, since the amount of scintillation light strongly depends on the type of interacting particle. Thanks to this two-channel approach, possible signal events (nuclear recoils) can be discriminated against background (as, in particular, electrons/gammas) by the ratio of light to heat signal, referred to as light yield (LY). In such detectors even the recoiling target nucleus (sodium or iodine) may be identified, hence offering a very powerful tool to study rare nuclear processes.

We refer to the unit of NaI crystal and light detector as a \textit{detector module}. A basic schema of the module used for the measurement presented here is depicted in figure \ref{pic:module_schema}. The NaI crystal has the dimensions \unit[(30x30x20)]{mm$^{3}$} (\unit[66]{g}) and is not doped in contrast to the DAMA/LIBRA target crystal, which is operated at room temperature and, thus, needs a dopant (Tl at 0.1\% level) to have sufficient light output. 

The NaI is attached to a carrier crystal made of CdWO$_4$ (\unit[39.2]{mm} in diameter and \unit[1.6]{mm} in height) equipped with a Transition Edge temperature Sensor (TES). CdWO$_4$ was chosen as it is known to have good phonon propagation properties (relatively high Debye temperature, low heat capacity), is easy to handle for TES production and is a material with a good compromise regarding the phonon transmission probability from NaI to the carrier. 
The carrier crystal is mandatory as the thin film TES cannot be directly evaporated on NaI due to its hygroscopic nature and its low melting point. The interface between the carrier and the absorber crystal is made from a very thin layer of silicon oil which works as link for the phonons. The carrier is fixed to a copper frame using bronze clamps. The NaI crystal on the contrary is free-standing and only attached to the carrier crystal through the silicon oil interface. 

\begin{figure}[t]

  \begin{minipage}[t]{0.49\textwidth}
    \centering
    \includegraphics[height=5cm]{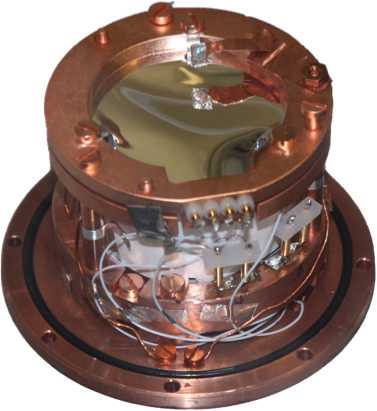}
    \caption{Closed detector module surrounded by a reflective foil to maximize light collection.}
    \label{fig:module}
  \end{minipage}%
  \hfill
   \begin{minipage}[t]{0.49\textwidth}
    \centering
    \includegraphics[height=5cm]{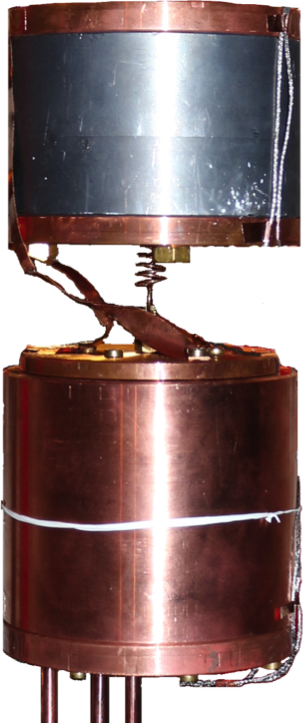}
    \caption{Air-tight copper container housing the detector module as mounted to the mixing chamber of the dilution refrigerator.}
    \label{fig:assembly}
  \end{minipage}
\end{figure}

The cryogenic light detector facing the NaI crystal consists of a sapphire wafer (\unit[40]{mm} in diameter and about \unit[460]{$\mu$m} in thickness) also equipped with a TES, but optimized for light measurement (conventional CRESST-II type).\footnote{The final COSINUS detector design, which is not realized for this measurement, will use a beaker-shaped light detector \cite{angloher_cosinus_2016} completely surrounding the target crystal.} To increase the absorption efficiency of the blue scintillation light a 1-$\mu$m-thick layer of silicon is epitaxially grown onto the sapphire disc (SOS = Silicon On Sapphire). The dedicated TESs on both detectors are based on superconducting tungsten thin films (\unit[200]{nm}, W-TES) which were produced at the Max-Planck-Institute for Physics in Munich. Each detector is paired with a dedicated heater used to stabilize the TESs in their desired operating point. The whole setup is surrounded by a reflective foil  (LuMirror\texttrademark~lateral side, VM2002\texttrademark~Radiant Mirror Film above light detector) as shown in figures \ref{pic:module_schema} and \ref{fig:module}.

The detector module is assembled in a glove-box under controlled atmosphere and finally housed in a copper container to avoid any contact of the NaI crystal with ambient, humid air at any time. To evacuate the container when mounted in the cryostat, we developed a dedicated cryogenic valve opening during the cool-down of the detector. Both the crystal and the light detector are exposed to a low-activity $^{55}$Fe X-ray source for energy calibration. 

\subsection{Experimental Setup and Data Acquisition}

The measurement was carried out in the test facility of the Max-Planck-Institute for Physics located at the deep underground site of LNGS. It is composed of a dilution refrigerator which allows detector operation at temperatures as low as \unit[7]{mK}. The cryostat is laterally surrounded by about \unit[20]{cm} of low-background lead to reduce the environmental $\beta / \gamma$-radioactivity.
The copper container is attached to the mixing chamber, the coldest point of the dilution unit, via a spring-loaded platform which mechanically decouples the detector from the cryogenic facility. The setup is a double stage pendulum, similar to the one described in \cite{Pirro}. A total of three stainless steel wires are used to connect the first stage copper plate to the mixing chamber, the second stage consists of a bronze spring and carries the detector platform. The TESs are read out with commercial dc-SQUID electronics (Applied Physics Systems company). The hardware-triggered signals are sampled in a \unit[328]{ms} window at a sampling rate of \unit[25]{kHz}. Both detectors are always read out simultaneously, independent of which one triggered. Detailed descriptions of the data acquisition, the control of detector stability and the pulse height evaluation and energy calibration procedures may (among others) be found in \cite{Angloher2009_run30, Angloher2005_cresstIIproof}.

\section{Results}
\label{sec:results}

In this paper we present data corresponding to a gross live time (before cuts) of \unit[303]{h}. The amplitude of each pulse is determined with a template fit. The templates (also called standard pulses) are obtained by averaging a large number of pulses of the same energy to extract a noise-free description of the pulse-shape. 

Two ingredients guarantee a precise energy calibration. Firstly, during the whole measurement the detectors were irradiated with $\gamma$-rays originating from an $^{241}$Am source, with an energy of \unit[59.541]{keV}. Secondly, we precisely measure the detector response function over the full dynamic range of the thermometers by injecting electrical pulses to the respective heaters. 

\subsection{Selection Criteria}

We apply several selection criteria to the events recorded. To determine the efficiency of each of those cuts we produce a data set of artificial  signal events. Creating these events is done by adding standard pulses - scaled in amplitude to map different desired energies - on empty baselines (randomly acquired noise samples), thus carrying also potential time-dependent noise changes. We apply the same energy reconstruction and selecting criteria on this data set, therefore the efficiency of a cut at a certain energy is given by the fraction of events surviving it.\footnote{This method was successfully used by CRESST, e.g.~in \cite{CRESST2,CRESST_LISE}.} Given the high pile-up rate in this measurement we want to stress that this method automatically accounts for the efficiency reduction due to pile-up events. This is the case since the empty baselines are acquired by reading out the detectors at a certain time regardless whether there was actually a pulse in the record window or not. 

\begin{figure}[t]
\centering
\includegraphics[width=0.5\textwidth]{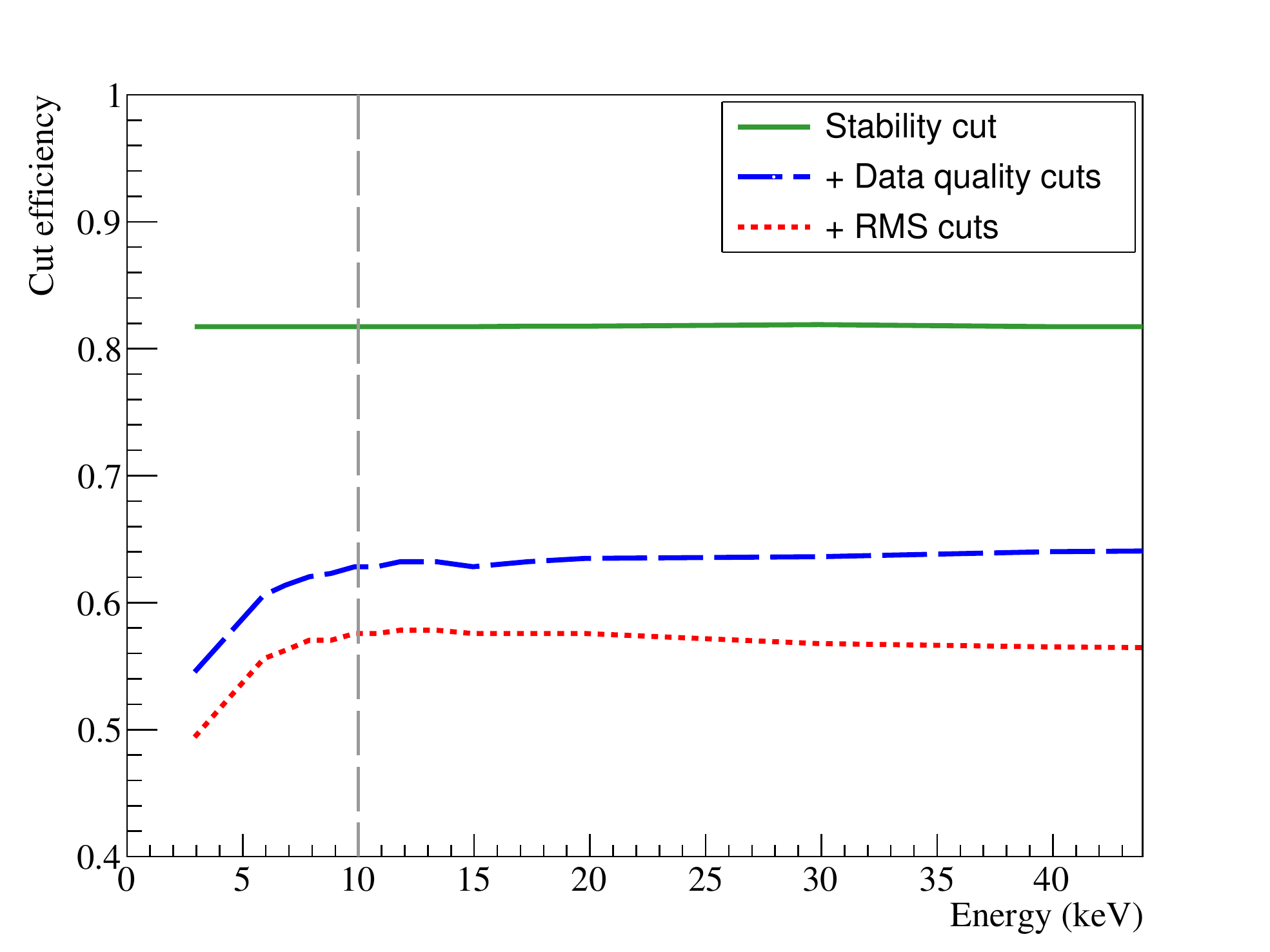}
\caption{Cut efficiencies (= survival probabilities for a valid event) after cumulative application of the selection criteria given in the legend. The gray dashed line marks the estimate for the hardware trigger threshold of 10~keV.}
\label{fig:cuteff}
\end{figure}

Firstly, we remove periods of unstable detector operations which are identified by the response of the detectors to injected (electrical) heater pulses being outside a predefined allowed range. The efficiency of this so-called stability cut is - as expected - completely independent of energy, as can be seen in figure \ref{fig:cuteff} showing efficiency of the stability cut as a function of energy in solid green.  

Secondly, we remove pile-up events, events with severely tilted baselines and finally events with a strong deviation from the nominal pulse-shape. The reason behind these cuts is to reject all events where a correct energy reconstruction might not be guaranteed; the efficiency after these cuts is drawn as dashed blue line in figure \ref{fig:cuteff}. For the same reason we apply a cut on the root mean square deviation (RMS) of the template fit which is generically sensitive to deviations from the nominal pulse-shape. The efficiency after this cut marks the final efficiency and is depicted as red dotted line in figure \ref{fig:cuteff}. In the region of interest between \unit[0-40]{keV} the cut efficiency is almost energy-independent and always above \unit[56]{\%} which corresponds to a net exposure of \unit[0.46]{kg days}. Although this value is already slightly higher than the  design goal specified in \cite{angloher_cosinus_2016} (\unit[50]{\%} in the energy range above \unit[2]{keV}) further improvements for stability, quality and RMS cuts are plausible when running the detector in a low-background environment with reduced pile-up rate.

\subsection{Pulse-shape}
The shape and temperature dependence of signals acquired from cryogenic detectors using TES thermometers is well modeled by the model for cryogenic particle detectors from F.~Pr\"obst et al.~\cite{Proebst}. In this detailed model a particle pulse is described as a superposition of two exponential pulses, given in equation~\eqref{eqn:proebst}. The first component is referred to as the non-thermal component since it accounts for the direct absorption of non-thermal phonons in the thermometer. Phonons that thermalize in the absorber are described by the second component.
Note, all particle interactions in the crystal, irrespective of whether through electron or nuclear recoils, first create high frequency phonons $\mathcal{O}$(100 GHz) with energies of a few \unit{meV}. These phonons are called \textit{non-thermal phonons} as thermal energies in the temperature range of the detector operation (approximately \unit[10]{mK}) are very small in comparison ($E\approx k_{B}T \approx$ \unit[1]{\textmu eV}).

The amplitudes of the two components are denoted with $A_{n}$ and $A_{t}$ for the non-thermal and the thermal component, respectively.\par

\begin{equation}
\Delta T_{e}(t)= \Theta (t) \Big[A_{n} \left( \mathrm{e}^{-t/\tau_{n}} - \mathrm{e}^{-t/\tau_{in}} \right) + 
A_{t} \left(\mathrm{e}^{-t/\tau_{t}}-\mathrm{e}^{-t/\tau_{n}}\right)\Big]
\label{eqn:proebst}
\end{equation}

The three different time constants present in equation~\eqref{eqn:proebst} are
\begin{itemize}
 \item[$\tau_{n}$:] the effective time constant for the thermalization of non-thermal phonons (life time of non-thermal phonons),
 \item[$\tau_{t}$:] the thermal relaxation time constant of the absorber temperature (influenced by the heat capacity of the absorber),
 \item[$\tau_{in}$:] the intrinsic thermal relaxation time constant of the thermometer (influenced by the coupling to the heat bath). 
\end{itemize}
Depending on the "strength" of the coupling of the thermometer to the heat bath, such detectors can be run in two different operating modes: calorimetric and bolometric. In the calorimetric mode the TES is only weakly coupled to the heat bath and, therefore, integrates the flux of non-thermal phonons. The latter are the dominant signal component at low temperatures, due to the very weak coupling of phonons and electrons. Consequently, in this configuration the signal height is ruled by the availability of non-thermal phonons and the heat capacity of the TES. On the contrary, a strong coupling of the TES to the heat bath results in a flux measurement of non-thermal phonons. In this bolometric operation, as also realized for the present NaI detector, the coupling of the thermometer rules the signal height. 

\begin{figure}
\centering
\subfigure[Two-component pulse model\label{fig:pulseA}]{\includegraphics[width=.49\textwidth]{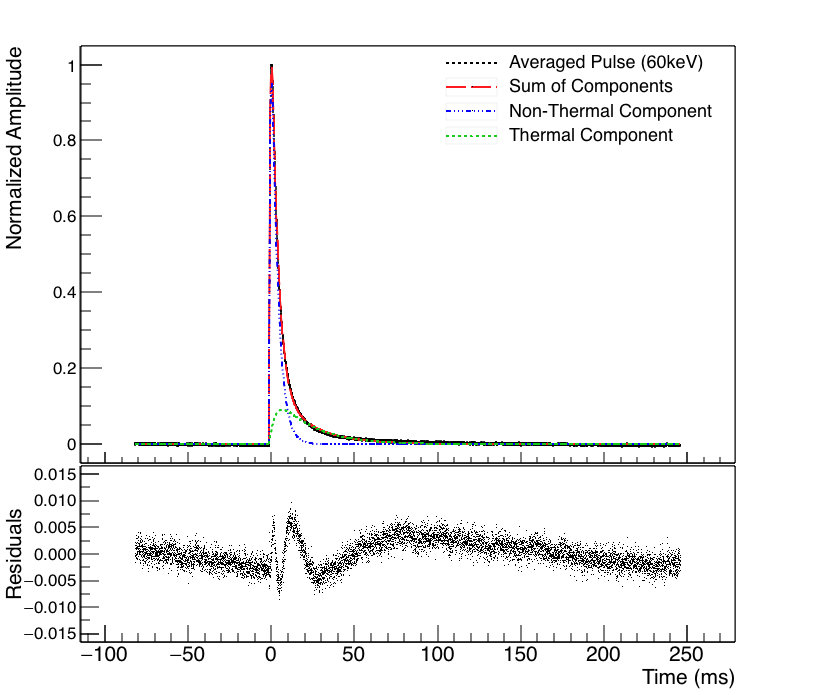}}
\subfigure[Three-component pulse model\label{fig:pulseB}]{\includegraphics[width=.49\textwidth]{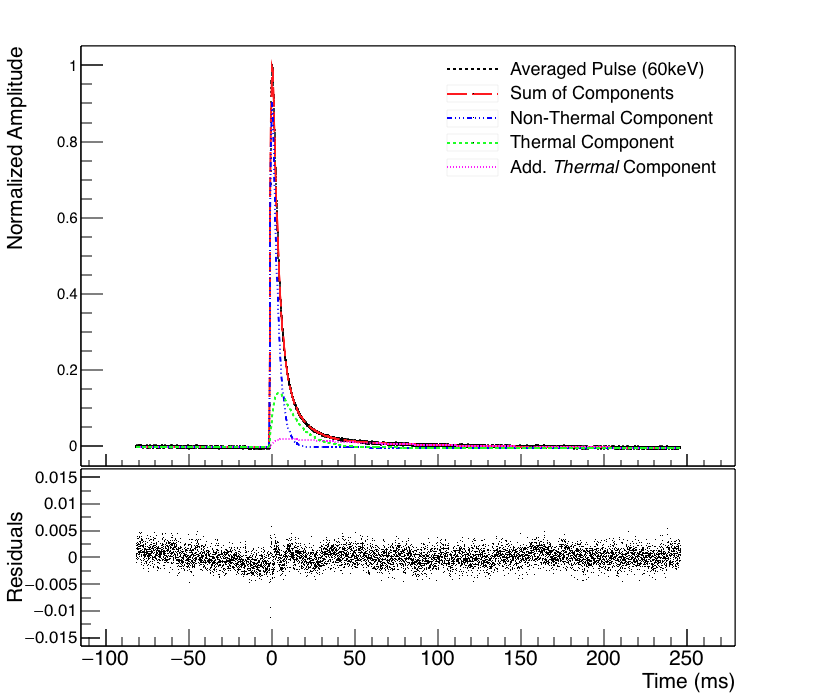}}
\caption{Upper windows: Fit of the standard event (dotted black), created by averaging a large number of 60~keV $\gamma$-events ($^{241}$Am calibration), with two (a, left) and three exponential components (b, right); see legend for description of components. The lower windows show the residuals of the respective fits. Resulting fit parameters are listed in table \ref{tab:PulseFitParams}. } 
\label{fig:pulse}
\end{figure}

In figure \ref{fig:pulseA} we show a fit (together with the residuals) of a standard pulse from the NaI detector using the model of F.~Pr\"obst et al.~\cite{Proebst}. The standard pulse was built by averaging a large number of $\gamma$-events from the $^{241}$Am-calibration source. The two components of the fit model \ref{fig:pulseA} do not allow for a perfect description of the particle pulse. 
This is in particular interesting since other materials including Al$_2$O$_3$, CaWO$_4$, ZnWO$_4$, CdWO$_4$, CaF, and TeO$_2$ measured with the same experimental technique did not show such behavior but are perfectly described by the model of Pr\"obst et al.. Moreover, the fitted NaI pulse looks like a compromise trying to marry the fast rise of the pulse with the quite longish thermal decay constant. By adding a third thermal component to the model the fit result can be significantly improved as can be seen from the residuals in \ref{fig:pulseB}. 
Table \ref{tab:PulseFitParams} lists the parameters as determined by the fits. Three parameters are fit in addition to equation ~\eqref{eqn:proebst}: \textit{Bl. level} and \textit{Bl. tilt} allow for a linear approximation of the baseline, whereas t$_0$ accounts for the onset of the pulse in the record window. 
In order to legitimize the introduction of a second thermal component further studies with an improved detector (signal-to-noise ratio) and an extended record window to fully cover the pulse decaying back to baseline are foreseen. This is mandatory to arrive at a physical description of the observed pulse-shape.

\subsection{Detector Performance}

\begin{figure}[t]
\centering
\begin{minipage}[t]{0.49\textwidth}
\centering
\includegraphics[width=\textwidth]{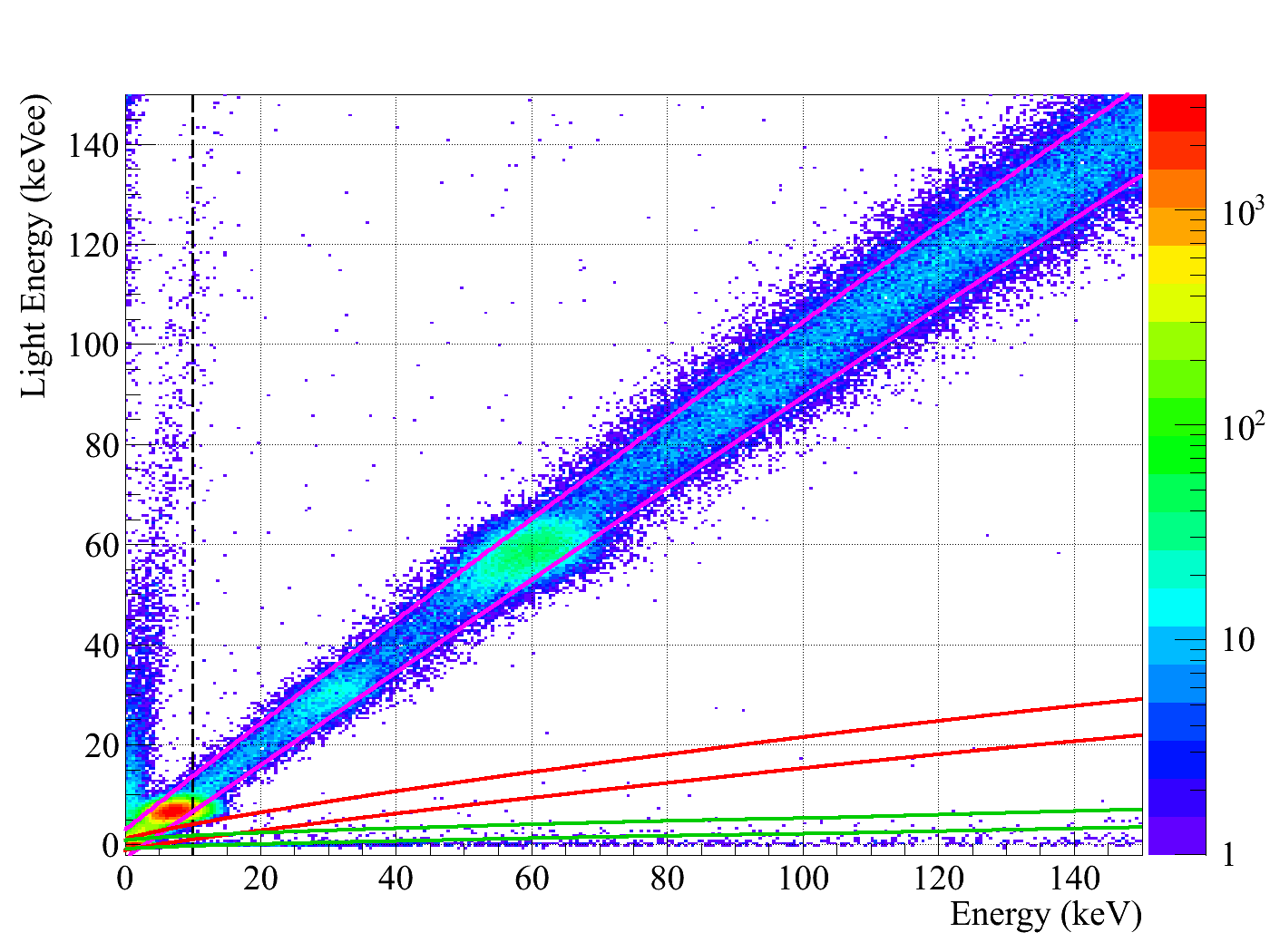}
\caption{Data corresponding to a net exposure of \unit[0.46]{kg days} in the light vs.~(phonon) energy plane. The magenta solid lines confine the central \unit[80]{\%} of the $\beta / \gamma$-band. Accordingly, the red and green bands correspond to recoils off sodium and iodine, respectively. The black dashed lines indicates the estimate of the trigger threshold of the phonon detector at 10~keV.}
\label{fig:LY}
\end{minipage}%
\hfill
\begin{minipage}[t]{0.49\textwidth}
  \includegraphics[width=\textwidth]{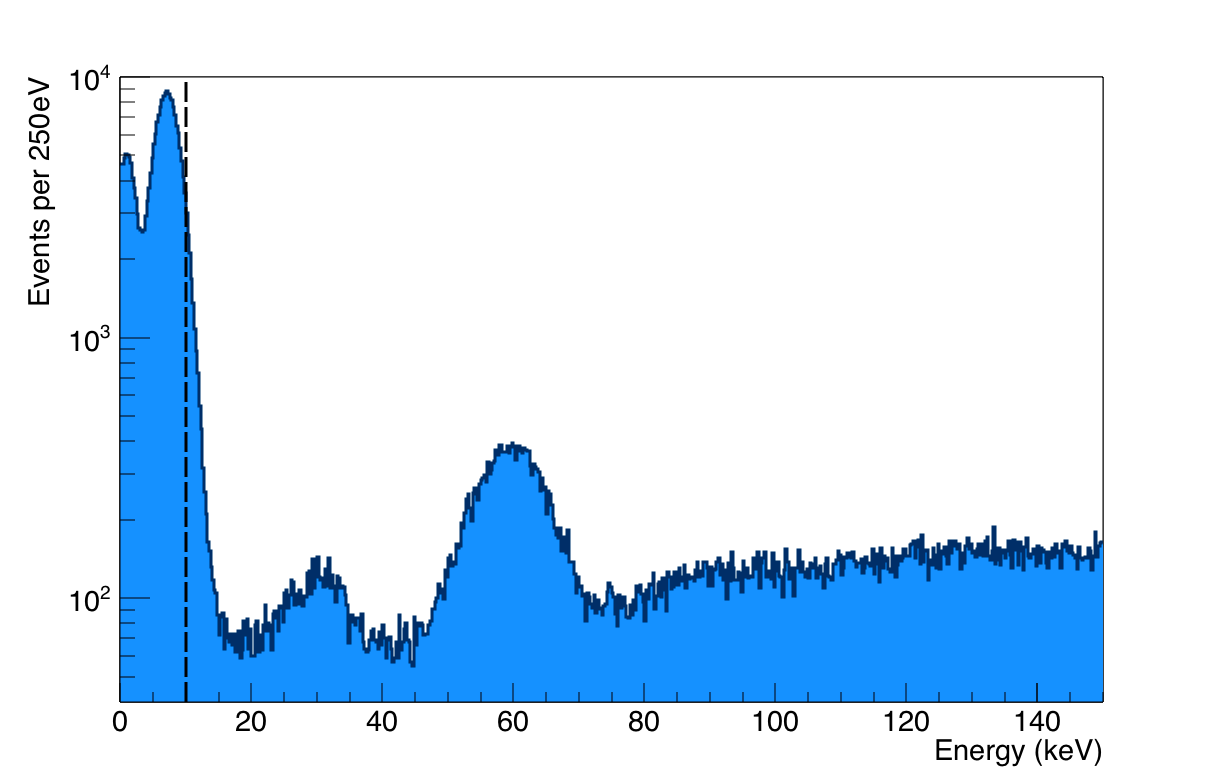}
  \caption{(Phonon) energy spectrum showing the $^{241}$Am $\gamma$-peak at 60~keV, a superposition of iodine escape peaks at 30~keV and the X-rays from an $^{55}$Fe source at $\sim$6~keV. The estimate of the trigger threshold of the phonon detector is illustrated with a black dashed line at 10~keV.}
  \label{fig:EnergySpectrum}
\end{minipage}
\end{figure}

Figure \ref{fig:LY} shows a 2D-histogram of the events surviving all cuts in the light vs.~phonon energy plane. The dominant population of events along the diagonal originates from $\beta / \gamma$-events. We fit the corresponding band as e.g.~described in \cite{StraussQF}. The mean of the band accounts for a linear dependence of the scintillation light output on the energy deposited in the NaI crystal. The width of the band at a certain energy is modeled by a Gaussian function centered at the mean of the band and includes contributions from the baseline resolutions of phonon and light detector, from Poisson statistics scaling with the square-root of the number of photons detected in the light detector and from potential position dependencies of the light detection. The outcome is the two magenta solid lines marking the central \unit[80]{\%}  $\beta / \gamma$-band. Based on this band and the energy-dependent quenching factors \footnote{For this application the quenching factor quantifies the reduction in light output of a certain event type compared to a $\beta / \gamma$-event of the same energy.} \cite{Tretyak} we then calculate the bands corresponding to recoils off sodium (red) and iodine (green). However, since the quenching factors in \cite{Tretyak} were determined at room-temperature for Tl-doped NaI crystals, a dedicated low-temperature measurement with undoped NaI is mandatory. We foresee to perform this measurement within the initial COSINUS R\&D phase using a low-temperature neutron scattering facility (previously used by the CRESST collaboration \cite{StraussQF}). The black dashed line at \unit[10]{keV} marks the estimated hardware trigger threshold of the phonon detector. 

The almost horizontal event population for very small light energies (below the green iodine recoil band) originates from particle events taking place in the CdWO$_4$ carrier crystal (see figure \ref{pic:module_schema}). The phonon signals of such carrier events are roughly 50 times larger than those of NaI crystal events, consequently causing a significant overestimation of the phonon energy. This overestimate results in the very small slope of this event population in the light vs.~energy plane. This interpretation is supported by expected pulse-shape differences in the phonon and light signal observed at higher energies. However, for energies near threshold these differences are not pronounced enough, thus no cut on these pulse-shape differences was applied. We foresee two measures to avoid the leakage of this background into the nuclear recoil bands (the ROI) in the future: Firstly, a lower count rate in a low-background facility will allow us to increase the record length, thus boosting pulse-shape discrimination power by exploiting the differences in the decay of the pulses. Secondly, the final beaker-shaped light detector will result in an increased light detector performance which in turn increases the separation of the \textit{carrier band} and the nuclear recoil bands in the light yield-energy plane.

Three prominent structures become evident in the $\beta$/$\gamma$-band, also clearly visible in the phonon energy spectrum shown in figure \ref{fig:EnergySpectrum}. The peak at \unit[60]{keV} is attributed to $\gamma$-rays from $^{241}$Am, the peak at \unit[30]{keV} is a superposition of several iodine escape peaks, which was already visible in a previous CsI measurement using the same experimental setup \cite{CsI_Tech}. Although being below the threshold of the phonon detector (\unit[10]{keV}), a considerable fraction of X-rays with energies of \unit[5.89]{keV} and \unit[6.49]{keV} ($^{55}$Mn K${_{\alpha,\beta}}$) from the $^{55}$Fe-source are triggered by the light detector (threshold: \unit[$\sim$1]{keVee} \footnote{ee=electron equivalent}) which makes up for the peak below (phonon) threshold. 
The remaining background \textit{rising} towards \unit[150]{keV} originates from environmental $\gamma$-radiation in the Gran Sasso underground laboratory \cite{malczewski_gamma_2013}.
The hardware trigger threshold of roughly \unit[10]{keV} was balanced between a setting as low as possible versus a setting high enough not to exceed the maximal tolerable overall rate for stable detector operation. Two aspects indicate that a lower hardware trigger threshold would have been achievable with this detector module in a low-background environment. Firstly, the energy reconstruction still works for sub-threshold energies, see the X-ray peak at around \unit[6]{keV}. Secondly, experience from CRESST \cite{CRESST_LISE} shows that a threshold setting of five times the baseline resolution is feasible, which for the given detector would yield a value of: $\sim 5\cdot \sigma_{E=0} = \sim \unit[5.6]{keV}$ (see next paragraph). 

The data show a strong dependence of the phonon energy resolution as a function of deposited energy, in particular in the light of experience from previous measurements of e.g.~CaWO$_4$ crystals. While  the energy  resolution of electrical pulses injected to the heater, which is a measure of the resolution at zero energy,  is $\sigma_{E=0}$=\unit[$(1.12 \pm 0.01$(stat.))]{keV} the  width of the $^{241}$Am peak  is  $\sigma$=\unit[$(5.02 \pm 0.06$(stat.))]{keV}. The two other peaks discussed above yield values of  $\sigma$=\unit[2.1]{keV}@\unit[$\sim$6]{keV} ($^{55}$Fe) and  $\sigma$=\unit[3.7]{keV}@\unit[$\sim$30]{keV} (I-escape). However, as  those two features are not single peaks their widths are rather an upper limit on the resolution, than  a  precise determination of the latter.

\subsection{Amount of Measured Light} \label{subsec:AbsoluteLY}

We define the absolute light yield as the fraction of scintillation light measured in the light detector to the total deposited energy by a $\beta / \gamma$-particle  interaction in the NaI crystal. To determine this quantity we compare the amplitude of a scintillation light pulse induced from a \unit[59.541]{keV} $\gamma$-particle - originating from the $^{241}$Am calibration source - to a direct hit of the light detector from the $^{55}$Fe X-ray calibration source. To account for the pulse-shape differences between a direct hit and a scintillation light event the amplitudes are weighted by the integrals of the respective standard pulses \cite{kiefer_-situ_2016}. We find that \unit[3.7]{\%} of the energy deposited in the NaI crystal is measured by the light detector. This value is already close to the final COSINUS design goal of \unit[4]{\%} \cite{angloher_cosinus_2016} which should easily be reachable with the beaker-shaped light detectors foreseen for the final design thanks to their significantly enhanced light collection efficiency. Given the mean energy of \unit[3.3]{eV} for a scintillation photon of undoped NaI, we measure an average number of 11.2 photons per keV of energy deposition (for a  $\beta / \gamma$-event).

\begin{table}[t]
\centering

\begin{tabular}{ l c c }
  Parameter & 2 components & 3 components \\
   \hline
  Bl. level           & $(-7.6\pm0.5)\cdot10^{-4}$      &  $(-2.09\pm0.03)\cdot10^{-3}$ \\
  Bl. tilt (ms$^{-1}$)& $(-2.9\pm0.3)\cdot10^{-6}$     &  $(-1.16\pm0.04)\cdot10^{-5}$\\
  t$_0$ (ms)          & $-1.292\pm0.0006$             &  $-1.299\pm0.0004$\\
  A$_\text{n}$        & $1.669\pm0.003$              &  $1.828\pm0.004$   \\
  A$_\text{t}$        & $0.161\pm0.001$              &  $0.341\pm0.009$   \\
  A$_\text{t2}$       & -                            &  $0.026\pm0.002$   \\
  t$_\text{in}$ (ms)  & $0.656\pm0.002$              &  $0.711\pm0.001$ \\
  t$_\text{n}$  (ms)  & $3.81\pm0.01$                &  $3.12\pm0.01$   \\
  t$_\text{t}$  (ms)  & $21.1\pm0.1$                 &  $10.5\pm0.1$    \\
  t$_\text{t2}$ (ms)  & -                            &  $75.1\pm1.9$    \\
\end{tabular}
\caption{Parameters  determined by the fit of the pulse model (see equation \ref{eqn:proebst}) to the average pulse (see figure \ref{fig:pulse}) including fit uncertainties.  Three parameters are fit in addition to equation \ref{eqn:proebst}, \textit{Bl. level} \& \textit{Bl. tilt} describing a linear approximation of the baseline and t$_0$ accounting for the onset of the pulse in the record window.} 
\label{tab:PulseFitParams}
\end{table}

\section{Conclusion and Outlook}
In this manuscript we, to our knowledge, report about the first successful measurement of a NaI crystal as cryogenic detector measuring simultaneously the phonon and light signal from a particle interaction. We observe a linear relation between light output and the energy deposited in the NaI crystal. The achieved energy threshold for our prototype is around \unit[10]{keV}, which is still well above the COSINUS design goal of \unit[1]{keV}. The energy resolution at the \unit[60]{keV} $\gamma$-calibration-line is \unit[5.0]{keV} but showing a significant improvement towards lower energies. 
The key-issues to be addressed in order to  meet the performance goals of the COSINUS project \cite{angloher_cosinus_2016} include the development of specifically designed sensitive TES thermometers and further optimization of the interface between the sodium iodide and the carrier crystal. In the presented detector setup we observe 3.7\% of the deposited energy in form of scintillation light. Given the enhanced light collection efficiency expected for the beaker-shaped silicon light absorber foreseen for the final COSINUS design, the project goal of 4\% most probably will be exceeded. 

The COSINUS detector is the first of its kind on the way to a NaI-based detector with low nuclear recoil threshold and particle discrimination, two truly unique advantages in the field of experiments with NaI target. Quintessential the successful operation of the prototype module was the first and essential step to converge to the COSINUS goal. Once reached, a moderate exposure of $\mathcal{O}$(\unit[100]{kg-days}) will be sufficient to confirm or rule-out a nuclear recoil origin of the DAMA/LIBRA dark matter claim, independent of the dark matter halo and the dark matter nucleus interaction mechanism.

\section*{Acknowledgements}
This work was carried out in the frame of the COSINUS R\&D project funded by the  Istituto Nazionale di Fisica Nucleare (INFN) in the Commissione Scientifica Nazionale 5 (CSN5). In particular, we want to thank the LNGS mechanical workshop team E.~Tatananni, A.~Rotilio, A.~Corsi, and B.~Romualdi for continuous help in the overall setup construction and M.~Guetti for his cryogenic expertise and his constant support. We are also very grateful to the people which helped us to keep the dilution refrigerator cold, in particular M.~W\"ustrich and J.~Rothe. Furthermore, we want to thank LNGS for the hospitality and the onsite support. \\
We express our deep appreciation to Dr. Wolfgang Seidel, the pioneer in the field of low temperature detectors whose contribution, also to this work, was of immense impact.
\bibliographystyle{JHEP.bst}

\bibliography{main}

\end{document}